\documentclass{aastex631}

\newcommand {\ci} {\ion{C}{1}}
\accepted{November 29, 2022}

\submitjournal{ApJL}

\graphicspath{{./}{figures/}}

\begin{document}

\shorttitle{CI kinematic structure in HD 163296}
\shortauthors{Alarc\'on et al.}

\title{A localized kinematic structure detected in atomic carbon emission spatially coincident with a proposed protoplanet in the HD 163296 disk}

\correspondingauthor{Felipe Alarc\'on}
\email{falarcon@umich.edu}

\author[0000-0002-2692-7862]{Felipe Alarc\'on }
\affiliation{Department of Astronomy, University of Michigan,
323 West Hall, 1085 S University, Ave.,
Ann Arbor, MI 48109, USA}

\author[0000-0003-4179-6394]{Edwin A. Bergin}
\affiliation{Department of Astronomy, University of Michigan,
323 West Hall, 1085 S University, Ave.,
Ann Arbor, MI 48109, USA}

\author[0000-0003-1534-5186]{Richard Teague}
\affiliation{Department of Earth, Atmospheric, and Planetary Sciences, Massachusetts Institute of Technology, Cambridge, MA 02139, USA}
\affiliation{Center for Astrophysics | Harvard \& Smithsonian, 60 Garden Street, Cambridge, MA 02138, USA}



\begin{abstract}

Over the last five years, studies of the kinematics in protoplanetary disks have led to the discovery of new protoplanet candidates and several structures linked to possible planet-disk interactions. We detect a localized kinematic bipolar structure in the HD 163296 disk present inside the deepest dust gap at 48 au from atomic carbon line emission. HD 163296's stellar jet and molecular winds have been described in detail in the literature; however, the kinematic anomaly in \ci \ emission is not associated with either of them. Further, the velocity of the kinematic structure points indicates a component fast enough to differentiate it from the Keplerian profile of the disk; and its atomic nature hints at a localized UV source strong enough to dissociate CO and launch a \ci\ outflow, or a strong polar flow from the upper layers of the disk. By discarding the stellar jet and previously observed molecular winds, we explore different sources for this kinematic feature in \ci \ emission that could be associated with a protoplanet inflow/outflow, or disk winds.

\end{abstract}



\section{Introduction} 

Over the past few years, the analysis of gas kinematic features traced by the emission of millimetric rotational lines has been proposed as a useful method to detect putative forming-planets in circumstellar disks \citep{Pinte..2022}. Embedded planets in protoplanetary disk generate kinematic footprints in the Keplerian pattern traced by molecular emission \citep{Perez..Seba}. Such footprints, either in pressure profiles or velocity deviations, have been observed in a handful of cases, thanks to the Atacama Large Millimeter and sub-millimeter Array (ALMA), proving the usefulness of the method \citep{Teague..2018,Pinte..2018, Pinte..2022, Izquierdo}. 

Much of the focus on disk kinematics have been related to the observation of planetary wakes. However, there are other possible kinematic features due to planets that could be observed.
 This includes meridional flows, and the inflow of material towards the accreting protoplanet \citep{Fung...Chiang, Rabago}.    Meridional flows and surface infall has been isolated in HD 163296 and HD 169142 \citep{Teague..2019, Yu..et..al..21}, but other protoplanet-induced features remain undetected at present. There is the possibility of molecular outflows and/or jets powered by the circumplanetary accretion disk and its internal angular momentum transport \citep{Pudritz..et..al..2007}. Such outflows linked to accreting protoplanets have been proposed in the literature, although without observed detections so far \citep{Wolk..Beck..1990, quillen..trillin..98,Fendt..2003, Lubow..Martin, Gressel..et..al..2013}.

HD 163296 disk stands out as one of the most comprehensively studied disks to date, having been extensively observed and modeled in detail. It has a rich dust millimeter continuum structure showing gaps, rings and a dust crescent that has been associated with an unseen giant planet \citep{Isella..2018}. Several models have proposed that massive planets are responsible for carving such structures \citep{liu..et..al..2018, Zhang..DSHARP}. Moreover, studies of its gas kinematics structure have been able to infer the presence of planets through Keplerian deviations \citep{Teague..2018, Teague..MAPS, Pinte..2018, Izquierdo..2022, Calcino..2022}. Thus, the HD 163296 disk is one of the best candidates to study planet formation in action.

In this paper, we present the detection of a strong kinematic deviation in the HD 163296 disk traced by \ci\ line emission which has not been observed in other molecular tracers \citep{Thi..2004, MAPS_Charles, Teague..MAPS, Vivi..MAPS}. We argue that such a kinematic feature can be tentatively associated with a protoplanet candidate or a disk wind that has not been observed before. We organize the present paper as follows. In Section \ref{sec:methods} we describe the datasets used from the ALMA Archive and the methodology used for the gas kinematics analysis. We present the results of our data analysis in Section \ref{sec:results}. In Section \ref{sec:disc} we discuss the possible sources of the kinematic structure observed in \ci\ emission. Finally, we summarize the findings of this paper in Sect. \ref{sec:summary}


\begin{figure}[ht!]
\plotone{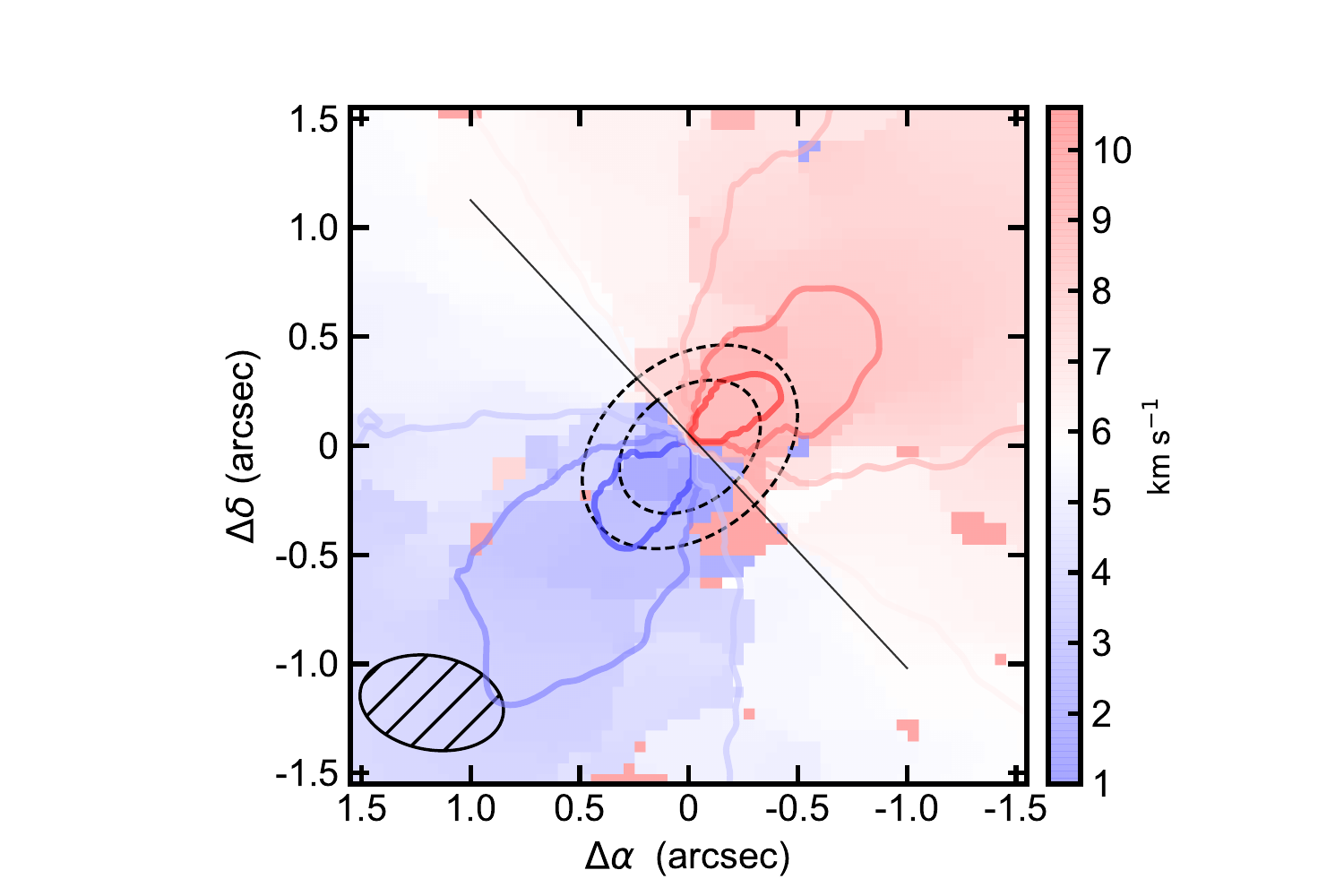}
\caption{Velocity centroid map of the \ci\ $^3$P$_1$-$^3$P$_0$ line emission using \texttt{bettermoments} with a Gaussian fit. Overlaid are the contours of the $^{13}$CO J=2-1 centroid map from the MAPS data \citep{MAPS_Charles} at 0\farcs 15 resolution. We show the region of the deepest gap in HD 163296 with dashed lines and the minor axis of the disk with a black line. The figure shows that the \ci\ line emission has a strong redshifted component,  $\sim$ 2 km s$^{-1}$ velocity offset,  located at $\sim$0\farcs 5 to the SW of the star, while the $^{13}$CO velocity centroid contours show a Keplerian profile. It is noteworthy that at that radius the deepest dust continuum gap of the HD 163296 disk is located \citep{Isella..2018}.
\label{fig:Mom1}}
\end{figure}

\begin{figure}[ht!]
\plotone{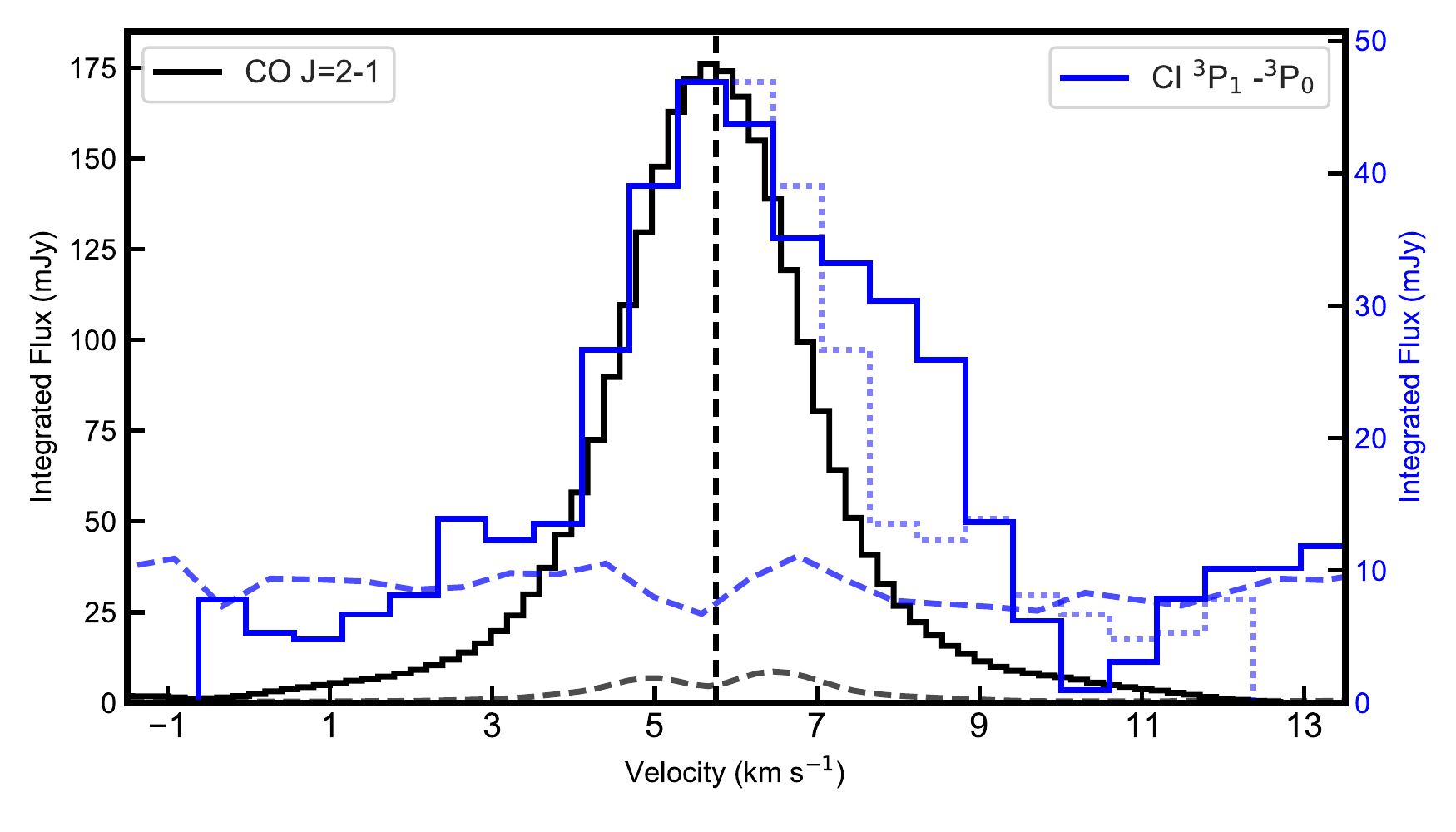}
\caption{Integrated spectra of the CO J=2-1 and the \ci\ $^3$P$_1$-$^3$P$_0$ line emission integrated over a 60 deg wedge, 60 au long, centered at 48 au on the SW orientation of the minor axis of the disk. The dim dashed lines show the respective noise levels of each integrated spectrum. Given that the integrated spectrum is centered on the minor axis,  a Keplerian disk should produce a symmetric structure around the stellar velocity, such as the CO J=2-1 emission; while the \ci\ line emission has an important redshifted component located in the deepest dust gap of HD 163296 at 48 au. We mirrored the blueshifted \ci\ emission with a dotted line to emphasize the redshifted emission excess in the wedge.
\label{fig:Spectrum}}
\end{figure}
.

\begin{figure}[ht!]
\plotone{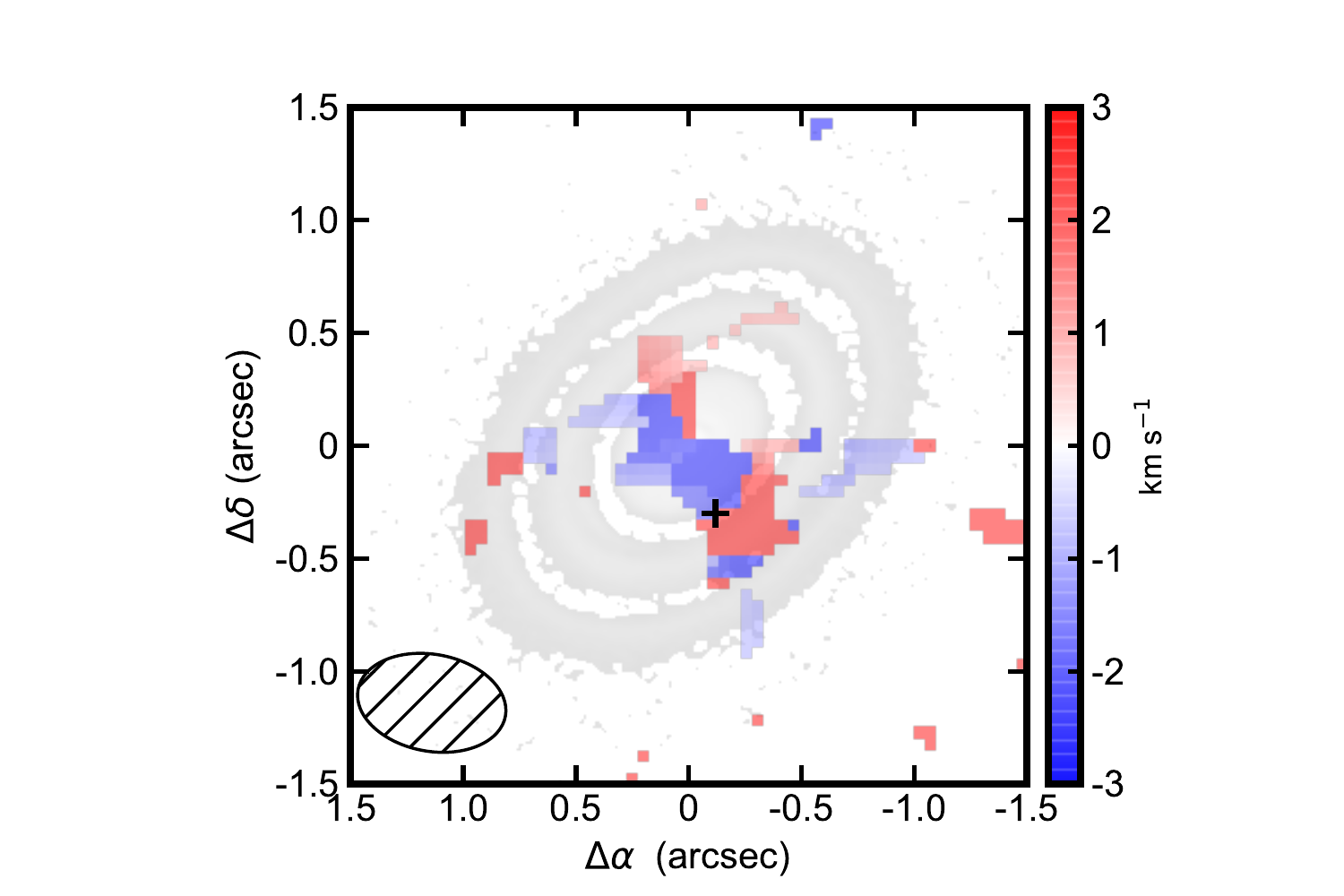}
\caption{Kinematic deviations between the velocity centroid map of the \ci\ $^3$P$_1$-$^3$P$_0$ line emission and the CO J=2-1 line from the MAPS survey \citep{MAPS} at matching spectral and spatial resolutions. The background is showing the DSHARP 1.3 mm dust continuum emission of the HD 163296 disk clipped at 65 $\mu$Jy beam$^{-1}$  \citep{DSHARP}. The CO J=2-1 velocity centroid was used as a proxy of the Keplerian rotational profile of the disk to isolate the kinematic anomaly observed in CI. The Figure shows a bipolar flow centered at the 48 au gap and at the same azimuthal location than the putative planet proposed by \cite{Isella..2018}.
\label{fig:Deviation}}
\end{figure}

\section{Observations and Data Reduction} \label{sec:methods}

We used archival ALMA observations of the HD 163296 disk from the \texttt{2013.1.00527.S} and \texttt{2015.1.01137.S} projects. The \texttt{2013.1.00527.S} project observed HD 163296 for 13 minutes with 37 antennas, while the \texttt{2015.1.01137.S} project targeted the HD 163296 disk with an integration time of 4 minutes.  We combined the datasets before imaging the \ci\ $^3$P$_1$-$^3$P$_0$ line observation in Band 8 with a rest frequency 492.161 GHz  using the CASA software \citep{CASA}. To properly combine them, both datasets were phase-centered in the celestial coordinates J2000 17\degr{}56$^{\prime}$21$^{\prime \prime}$ \  -21\degr{}57$^\prime$22.545$^{\prime \prime}$. The dataset was self-calibrated in two rounds with a solution interval of 30 s. The continuum was subtracted in the UV space with the \texttt{uvcontsub} routine and the images were reconstructed using the  \texttt{tclean} algorithm in the CASA software. Different Briggs weighting values were explored during the imaging process to find an optimal trade-off between sensitivity and spatial resolution. The final chosen weighting value was equal to 0.5. The channel spacing of the final cube was 600 m s$^{-1}$ to match the minimum independent spectral resolution of the initial datasets. The final RMS noise value was 33 mJy beam$^{-1}$ for a beam size of 0\farcs 65$\times$0\farcs 40 with a position angle of 79\fdg8 for each channel. Using those beam dimensions and signal-to-noise ratio, we obtained an astrometry precision $\sim$ 0\farcs19, equivalent to 20 au in the HD 163296 disk. 

After the datasets were combined, we analyzed the kinematics features in the \ci\ line emission using the \texttt{bettermoments} module \citep{bettermoments}, \texttt{eddy} \citep{eddy}, \texttt{disksurf} \citep{disksurf} and \texttt{GoFish} \citep{GoFish} python packages. By using those tools, we calculated the velocity-centroid maps and spectra from the disk using the \ci\ line emission. The velocity-centroid map in Figure \ref{fig:Mom1} was calculated by doing a Gaussian fitting inside the \texttt{bettermoments} module. For completeness, the velocity-centroid maps obtained from the Gaussian fit were compared to the ones from the quadratic fitting, showing the same localized kinematic feature. Given the spatial resolution of the observations, it was not possible to recover the emission height of the \ci\ line. Nevertheless, we were able to compare the \ci\  velocity-centroid map with the CO velocity-centroid from the MAPS data \citep{Teague..MAPS} by subtracting them.

\section{Results} \label{sec:results}

Figure \ref{fig:Mom1} shows the contours of the velocity-centroid maps of the $^{13}$CO J=2-1 line emission overlaid on the velocity-centroid map of the \ci\ line emission. In the Figure, the CO contours show the systemic rotational profile in the disk, while the \ci\ velocity-centroid map has a clear redshifted lobe located $\sim$ 50 au to the SW of the star. The redshifted component has a velocity offset $>$ 2 km s$^{-1}$, which suggests that is not linked to the Keplerian rotation of the disk. The redshifted lobe could be either a fast radially outwards component or vertical motion in the disk.

We provide the \ci\ and CO J=2-1 integrated spectrum of a 60\degr{} wedge with a 60 au radial extension centered in the gap at 48 au on the SW side of the minor axis of the disk in Fig. \ref{fig:Spectrum}. Given that the wedge is centered on the minor axis, a Keplerian disk should be symmetric around the systemic velocity of the star, 5.76 km s$^{-1}$ \citep{Teague..MAPS}. As expected, the CO spectrum is symmetric around the stellar systemic velocity, while the \ci\ emission has a bright redshifted component when compared with the blueshifted Keplerian component. This asymmetry hints at an extra velocity component in the disk which is reflected in the velocity-centroid maps shown in Fig. \ref{fig:Mom1}.

We show the subtraction of the \ci\ velocity-centroid map with the one from the CO J=2-1  MAPS data in Figure \ref{fig:Deviation} imaged at the same spatial and spectral resolution. By doing that we isolate the gas motion that is not associated with molecular tracers or the Keplerian disk. In the Figure, we also overlay the 1.3 millimeter dust continuum emission from the DSHARP survey \citep{DSHARP, Isella..2018} to show the location of the bipolar blob when compared with the dust continuum emission structure. In addition, the cross represents the proposed location of a putative planet as suggested by \cite{Isella..2018}.  A planet at this location would be able to create the dust continuum crescent observed within the gap at 48 au.  This putative planet appears to be spatially coincident with the \ci\ kinematic anomaly, both radially and azimuthally. There is an extra strong velocity deviation in the region close to the star. In that case, however, it is harder to isolate the emission from the inner disk given the beam size.

Finally, we note that this kinematic feature is not detected in other species as this is not found in the emission maps of any other species with resolved data.  For this purpose we searched the emission images of CO, C$_2$H, CN, HCO$^{+}$ and numerous other species/transitions observed in the MAPS program \citep{MAPS_Charles, Vivi..MAPS} for evidence of a similar anomaly.   No evidence is present in these deep datasets.  In additional support of its localized nature, no feature from HD 163296 at that velocity is observed in large-scale emission spectra from organic molecules with IRAM \citep{Thi..2004}.

\section{Discussion} \label{sec:disc}

Given its brightness, HD 163296 is one of the most well-characterized protoplanetary disks to date.   This system is associated with a well-characterized stellar jet that is also aligned along the NE-SW minor axis, coinciding  with the localized kinematic anomaly \citep{Xie}.   However, the velocities of the stellar jet have values ranging over 100 km s$^{-1}$; in contrast, the \ci\ feature is spread over only a couple  km s$^{-1}$. More importantly, the jet observed to the SW of the disk is blueshifted \citep{Gunther..2013, Ellerbroek, Xie, Kirwan}, which is expected given the orientation of the HD 163296 disk on the sky, i.e., the NE side being the closest to us \citep{Teague..MAPS}. On the contrary, the \ci\ feature at we observe at $\sim$50 au toward the SW is redshifted. At this location, such a feature could be interpreted as a vertical inflow/outflow, or a radial flow going outwards. A connection between the \ci\ kinematic feature and the slower molecular winds observed by \cite{Klaassen} and \cite{MAPS..Alice} is also unlikely as they are also blueshifted, with a velocity offset from the star $\sim$ -18 km s$^{-1}$.
At the location closest to the disk in the SW direction, the wind observed in CO emission is blue-shifted to $\sim$ -13 km~s$^{-1}$; excess \ci\ emission is at $\sim$8-10 km~s$^{-1}$.  Thus, the source of the \ci\ feature is kinematically decoupled from either the stellar jet or the  observed CO molecular wind given its velocity and its localized nature. 
Another potential origin for this feature could be the detection of a planetary wake.  However, the kinematic deviations associated with wakes are of order a few hundreds of m s$^{-1}$ for planets less massive than 1 Jupiter mass, and the Doppler flip would be azimuthally inverted to the one we observed \citep{Seba..2018}, i.e., negative and positive velocity deviations are azimuthally switched given the disk rotation.

The atomic nature of the kinematic feature requires a source of UV photons, strong enough to photodissociate CO. If the kinematic feature is tracing \ci\ flows from the upper layers of the disk, it requires a physical source to launch outflows at that speed that far from the star, which can be linked to planets. Conversely, if the feature traces \ci\ emission in the midplane, the UV photons need to be produced locally deep in the disk since CO self-shields UV photons coming from the star and the Interstellar Medium (see Appendix \ref{Appendix}). Such a source can be linked to a circumplanetary disk actively accreting in the midplane. We are not able to discard a stellar wind for the blue lobe, as it is closer to the star and also blueshifted; however, if the blueshifted lobe traces the disk wind, that still would not explain the source of the localized redshifted lobe at $\sim$ 50 au.

Given that the \ci\  velocity deviation is $>$ 2 km s$^{-1}$, i.e. $\delta v_0/v_K \sim 0.33$, for the anomaly to be associated with a planet, it has to be around 10 M$_{\rm Jup}$ for $\alpha \sim 10^{-3}$ (see Figure 7 in \cite{Rabago}). This estimate can be compared to previously reported values in the literature using observations of the disk at different wavelengths, e.g., \cite{Zhang..DSHARP} provide an estimate of a 2.18 M$_{\rm Jup}$ planet with $\alpha = 10^{-3}$ using dust millimeter continuum. Further, \cite{Mesa..2019} put an upper limit of 7 M$_{\rm Jup}$ for any possible companion with direct imaging observations.

\subsection{Potential Solution \#1: Protoplanet Infall}\label{disc1}

\begin{figure}[ht!]
\plotone{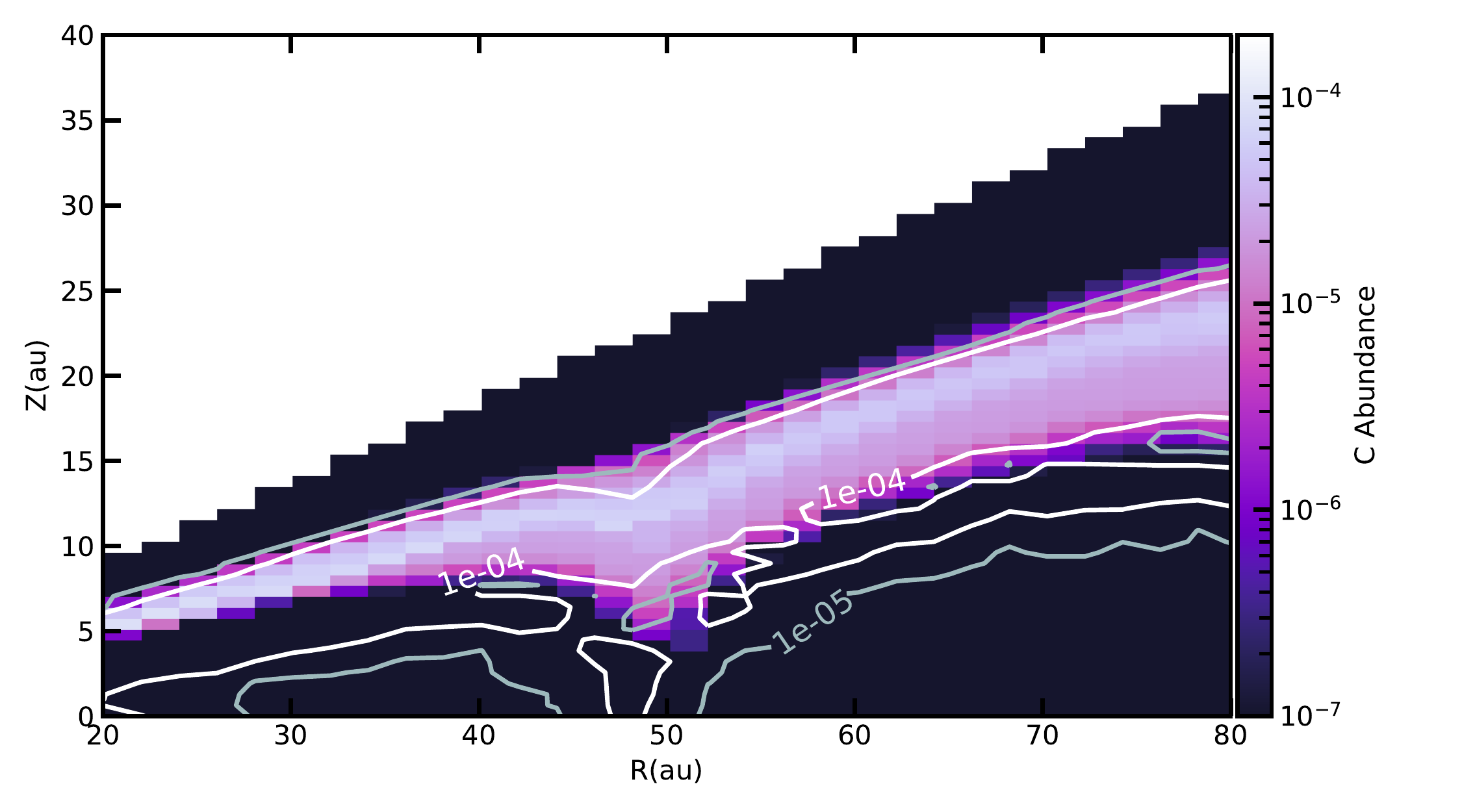}
\caption{\ci\ abundance in a thermochemical model of HD 163296 including the gaps and ring from dust continuum emission \citep{Isella..2018} ran with \texttt{rac2d} \citep{Du..2014}. Contours are drawn for the CO abundance in the disk. The Figure shows that inside the gap at 48 au, particularly towards the outer edge, the \ci\ abundance increases towards the midplane due to the transparency to UV in that region. Thus, it is possible that \ci\ line emission can trace vertical protoplanet infall in the disk.
\label{fig:CI_2D}}
\end{figure}

Vertical infall of material is expected in the presence of a massive planet, reaching levels even comparable to the local sound speed \citep{Fung...Chiang}. Tracing the infall through \ci\ line emission might be possible if it were dominated by disk surface gas directly exposed to stellar ultraviolet (UV) radiation, as the photodissociation of molecules becomes important. In the case of an embedded massive planet, it could carve a gap deep enough so the local pressure gradient dominates the gas flow even in the presence of disk winds \citep{Weber..2022}. Moreover, the depleted gap will increase the penetration of the stellar UV flux into the disk \citep{Facchini..2018, Alarcon..2020}. In that case, we expect the \ci\ abundance to increase in layers closer to the midplane, such that neutral atomic carbon might be a tracer of protoplanet infall. Figure \ref{fig:CI_2D} shows the \ci\ abundance in a thermochemical model of HD 163296 including its dust substructure. The Figure shows that \ci\ remains abundant at lower layers in the disk, closer to the midplane within the gap near 50~au. Moreover, the contours showing the CO abundance illustrate that as that region is more exposed to UV radiation, the photodissociation rate increases. If we assume the feature is tracing the surface of the disk, we can also put further constraints on the putative planet's mass by using the local sound speed in the disk. The sound speed in our thermochemical model at the \ci\ layer is $c_s = 1.5$ km s$^{-1}$, so the meridional speed of the \ci\ inflow agrees with the models of \cite{Fung...Chiang} for 1 to 4 Jupiter mass planets, i.e., $v_z \sim 1-2 c_s$, at the location of the planet.

Our model shows that \ci\ is clearly present in the gap, but the effects of photodissociation via stellar photons do not destroy enough CO in the midplane to raise the \ci\ abundance to values significant enough to be observable. From thermochemical models of HD 163296 with \texttt{rac2d} \citep{Du..Bergin..2014}, the predicted \ci\ emission for the $^3$P$_1$-$^3$P$_0$ line matching the beam size of the observations is 40 mJy beam$^{-1}$, which is not detectable given the noise value.

The peak intensity at the kinematic anomaly allows us to provide estimates of the vertical gas flow. By using the redshifted velocity, $v\sim$ 2 km s${-1}$ and the disk inclination, $i=$ 46.7 deg \citep{MAPS}, we can estimate a lower limit for the infall velocity, $v_z=v\cos {i} = 1.44$ km s$^{-1}$. Then, we can estimate the infall timescale using $t_{\mathrm{in}}=h/v_z$, where $h$, is the emission height of \ci. From Figure \ref{fig:CI_2D}, we use $h\approx 10$ au as a tentative \ci\ emission height, where most of the \ci\ is concentrated. Then, the estimated infall timescale is $t_{\mathrm{in}}=33$ yr. With the inferred \ci\ column density and the area of the redshifted anomaly we can estimate the mass accretion rate by using the following relationship:

\begin{equation}
    \dot{M} = m_{\rm H}A\frac{N_{CI}}{X_{CI}}\frac{1}{t_{\mathrm{in}}},
\end{equation}

\noindent where $A$ is the area of the redshifted anomaly, $m_H$ the hydrogen mass, $N_{CI}$ the inferred \ci\ column density, $X_{CI}$ the \ci\ abundance, and $t_{\mathrm{in}}$ the infall timescale. Using the peak intensity at the velocity anomaly, $I_{\nu} = 0.09$ Jy beam$^{-1}$, the inferred \ci\ column density is $N_{CI}=$ 3.5 x 10$^{17}$ cm$^{-2}$ for an excitation temperature T$_{\rm exc}$ = 45 K (see Appendix \ref{Appendix2}). Taking into account a transverse area, $A = \pi 20^2 $ au$^2$, and $X_{CI} = 10^{-5}$, the accretion rate is: $\dot{M} = 2 \times 10^{-4}\ M_{\rm Jup} \ \mathrm{yr}^{-1}$, which is within the expected range of accretion rates for growing planets from the literature \citep{Kley..2001, Szulagiy..2014, Tanigawa..2016}.

An additional point in favor of this scenario (and Solution \#2) is that the dust crescent is located in what would be the L5 Lagrangian point of the planet. It is expected that a massive planet, such as the one proposed to carve the dust gap at 48 au, can cause a dust pile-up in the L4 and L5 Lagrangian points \citep{Montesinos..et..al..2020}, which has also been observed in the LkCa 15 disk \citep{Long..et..al..2022}. Therefore, the azimuthal and radial location of the kinematic \ci\ feature agrees with the presence of multiple substructures in the disk.

\subsection{Potential Solution \#2: Protoplanet Outflow or Radial De-cretion}

If the \ci\ emission is arising from within deeper layers in the disk, there might be a local UV source generated by the protoplanetary accretion.  This would answer one puzzling aspect of this anomaly - why it is not seen in molecular emission (see \S~3). If a UV source is present to dissociate CO, then other species, which are more easily photodestroyed, would be as well. In the presence of optically thin C$^{18}$O emission, it is possible that a localized emission gap caused by CO photodissociation would be expected, although there will be an interplay between temperature and optical depth. However, the C$^{18}$O J=2-1 emission is still optically thick at 50 au in the HD 163296 disk, so it does not trace the midplane at 50 au.

If the \ci\ kinematic feature is not associated with infall onto a protoplanet, then it could be a localized source of outflow from an accreting protoplanet, and the kinematic feature is tracing an outflow behind the disk.  It is clear that protoplanets are born with accretion disks and emission from H$\alpha$ has been detected towards PDS 70 \citep{Haffert..2019}, which is associated with accretion onto the young protoplanets. Recently, a faint point source emission of H$\alpha$ has been observed in the HD 163296 disk \citep{Huelamo..et..al..2022} at the minor axis in the SW of the disk, although at a closer radius, $r=$ 171 mas. Given the presence of accretion disks, subsequent outflows or possible jets are expected in order to produce the loss of angular momentum \citep{Pudritz..PPV, pascucci..2020}. Given that the planet-disk interactions are strong enough, the \ci\ emission feature can also be tracing the radial de-cretion flows in the midplane towards the gap's edges \citep{Fung...Chiang, Rabago, Calcino..2022}.
 
Here the gas flow rate from 
 Section \ref{disc1} would reflect the outflow mass loss rate. Since we are tracing the backside of the disk, the diameter of the emitting area is constrained by the width of the dust gap, which is 20 au \citep{Jane..DSHARP}, so  we assume the area of the anomaly to be, $A=\pi$10$^2$ au$^2$. This emitting area is a factor of four smaller than the one assumed in our Potential Solution \#1, so the inferred accretion rate is also a factor of four smaller. 
 We can then use the relationship $\dot{M}_{\rm outflow} \approx 0.1 \dot{M}_{\rm acc}$ \citep{Pudritz..PPV}, leading to an estimated accretion rate of $\dot{M} = 2.5 \times 10^{-4}$ $M_{\rm Jup} \ \mathrm{yr}^{-1}$. Even though this value is still within the possible solutions for planetary accretion rates, it would be for a very strong accretor. This solution would also require the presence of a localized UV field to account for the lack of detection of a molecular counterpart.
 

\subsection{Potential Solution \#3: Disk Wind}

Another possibility is that the \ci\ feature is associated with disk winds. Disk winds are associated with the jets and magnetic activity of the central star or radiation-driven in the outer regions of disks. In detriment of this scenario, the stellar jet is blueshifted to the SW \citep{Xie}, which is the closest part of the disk. For the observed molecular wind in the disk, the velocity offset being at least a few tens of km s$^{-1}$ \citep{MAPS..Alice}. Additionally, theoretical and observational evidence have shown that disk winds associated with magnetic fields are launched from the inner disk, $<$ 10 au for Class II disks \citep{Goodson..1999, Matt..2005, pascucci..2020}. There is a possibility that the feature is tracing an FUV-driven wind on the far side of the disk, although it is expected that this would be observed as a diffused extended emission as well. Additionally, the bipolar nature of the feature offset from the central star makes this scenario less likely. Nevertheless, the blueshifted lobe is close to the stellar location, without additional high-resolution data we cannot discard that the blueshifted lobe is tracing a stellar wind. However, this scenario does not provide an origin for the offset of the bipolar structure and the localized redshifted lobe at $\sim$50 au.

Further, observations with the APEX telescope did not show strong \ci\ emission in HD 163296 \citep{Kama.et.al.2016}. This points out that the observed kinematic feature is local rather than a large-scale emission, which is usually the case for disk winds. Nevertheless, molecular winds launched from the disk surface at a few km s$^{-1}$ are still possible.

\section{Summary}\label{sec:summary}

In this work, we have presented the detection of a strong kinematic deviation in \ci\ line emission in the HD 163296 protoplanetary disk. We proposed that such structure is linked to either the presence of a strong UV source in the midplane of the disk, dissociating surrounding CO, which would explain the strong feature in \ci\ ; or the feature is coming from the disk atmosphere and is tracing inflow of gas into the midplane.  The fact that is not observed in CO or other molecular tracers, such as the ones observed by the MAPS program, suggests that this kinematic component is mostly atomic in nature and not molecular. Its atomic nature constrains the expected local physical and chemical conditions at the source of this feature.  The evidence for this feature is given below:

\begin{enumerate}
    \item It is located in the strongly depleted inner dust gap, which has been proposed to be carved by a massive Jovian planet.
    \item The projection of kinematic structure is parallel to the disk minor axis.
    \item The structure is observed in two independent datasets (see Appendix \ref{datasets}).
    \item The kinematic structure appears to be atomic in composition, requiring a localized UV source.
    \item The dust continuum crescent at the inner dust gap in HD 163296 is located in one of the dust pile-up Lagrangian points of the putative planet. 
    \item Planetary wakes from hydrodynamic simulations are at the level of a few hundreds of m s$^{-1}$. Both the molecular wind and the stellar jet's component to the SW of the disk are blueshifted, while the strongest emission in the \ci\ kinematic structure is redshifted. The velocity of the molecular wind in HD 163296 ranges between -20 and -18 km s$^{-1}$ offset from the stellar velocity \citep{MAPS..Alice}, while the blue-shifted component of the stellar jet has a velocity, $v_{\rm jet} = 280$ km s$^{-1}$ \citep{Xie}. Therefore, we discard other sources of velocity structure.
\end{enumerate}

Deeper observations of \ci\ $^3$P$_1$-$^3$P$_0$ line emission at higher spatial resolution and precision are needed to confirm the link between the \ci\ kinematic feature in the HD 163296 disk and a putative protoplanet. If a protoplanet nature is confirmed, the \ci\ kinematic feature would be the first detection of an atomic carbon protoplanetary footprint, introducing a novel technique of protoplanet detection, while providing further constraints in the planet formation process and composition of the planet-feeding gas. If the source of the \ci\ feature is coming from either a disk wind or planetary accretion, both of which require a local UV source, it would produce a distinctive feature in atomic forbidden lines, some of which are potentially detectable with JWST. Additionally, a more robust velocity-centroid map of the HD 163296 disk in \ci \ emission with ALMA would provide further evidence to disentangle the different possible scenarios.

\begin{acknowledgments}

This paper makes use of the following ALMA data:  ADS/JAO.ALMA\#2013.1.00527.S. , ADS/JAO.ALMA\#2015.1.01137.S.   ALMA is a partnership of ESO (representing its member states), NSF (USA) and NINS (Japan), together with NRC (Canada), MOST and ASIAA (Taiwan), and KASI (Republic of Korea), in cooperation with the Republic of Chile. The Joint ALMA Observatory is operated by ESO, AUI/NRAO and NAOJ. The National Radio Astronomy Observatory is a facility of the National Science Foundation operated under cooperative agreement by Associated Universities, Inc.
The National Radio Astronomy Observatory is a facility of the National Science
 Foundation operated under cooperative agreement by Associated Universities, Inc.

 We thank the anonymous referee for the insightful reading of the manuscript, and for providing useful and helpful comments to improve the quality of this work.

\end{acknowledgments}

%

\facilities{ALMA}


\software{astropy, \citep{2013A&A...558A..33A,2018AJ....156..123A},  
          CASA \citep{CASA},
          bettermoments \citep{bettermoments},
          eddy \citep{eddy},
          disksurf \citep{disksurf},
          GoFish \citep{GoFish}
          }



\appendix


\section{CO Self-Shielding}\label{Appendix}

\begin{figure}[ht!]
\plotone{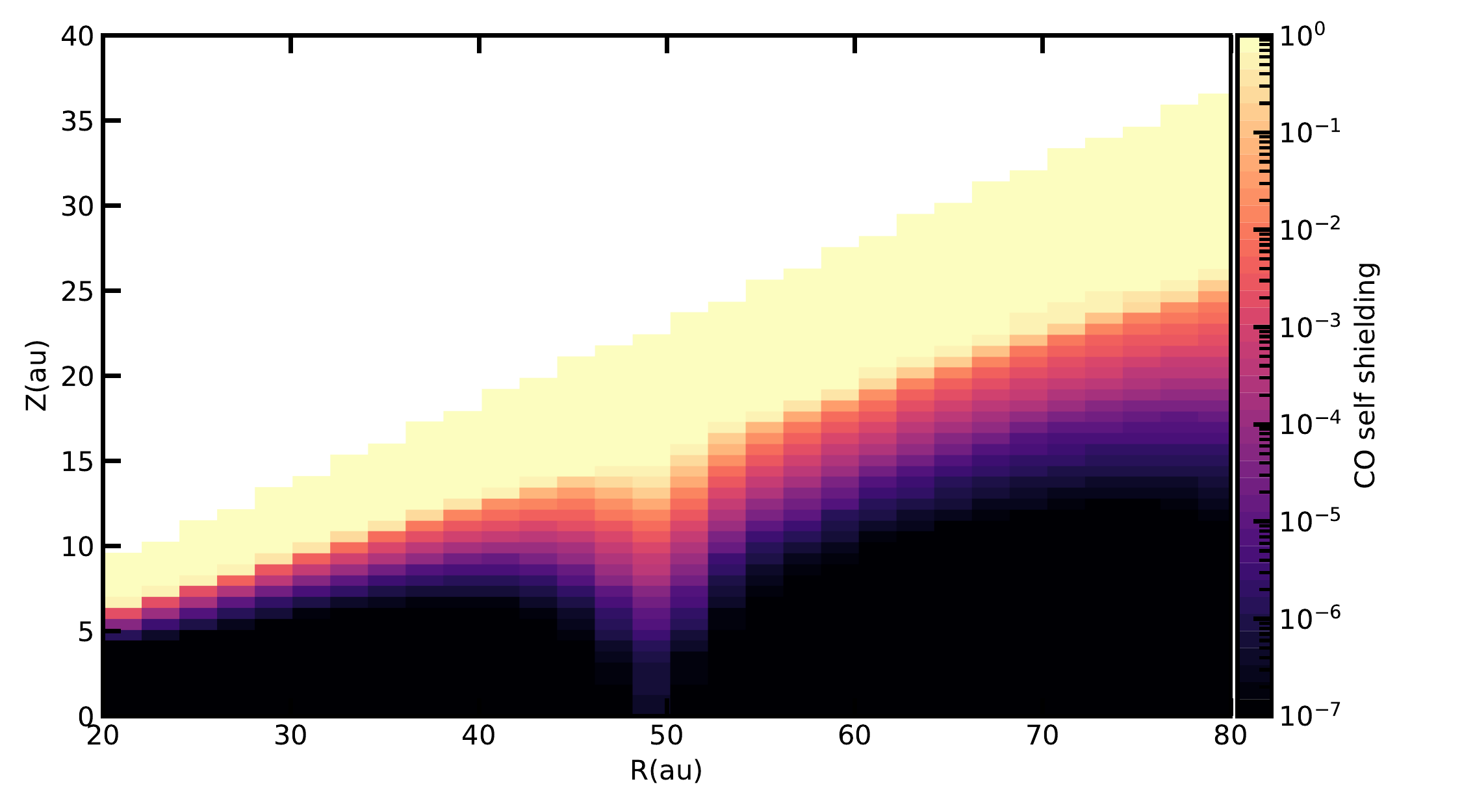}
\caption{CO self-shielding for the HD 163296 disk model with a dust-depleted gap at 48 au. CO self-shields itself even inside the gap. However, the content depletion inside the gap allows \ci\ formation through CO destruction at the deeper layer. The model shows that CO self-shields $\sim$ 99\% of the CO photodissociation photons 10 au above the midplane at 48 au. 
\label{fig:CO_SS}}
\end{figure}

One of the proposed solutions for the \ci\ kinematic feature is due to protoplanet infall from the upper layers of the disk. However, since the feature is not observed in molecular tracers, it requires that the inflow is actually coming from the very top layers of the disk, or the atomic layer of the disk is pushed down due to dust depletion. Another important factor to consider, in particular for \ci\, is CO self-shielding \citep{Visser..et..al..2009}. As CO self-shields itself it avoids its own destruction due to high-energy photons. We show the CO self-shielding factor in Figure \ref{fig:CO_SS}. It shows that in a 99\% dust-depleted gap, the CO self-shielding layer is pushed down. However, that push is not significant enough to put the \ci\ layer inside the Hill's sphere of a 3 Jupiter mass planet at that location. if the inflow scenario is the source of the kinematic asymmetry, it points to a chemical mixing between the upper layers of the disk and planet-feeding gas. If the source of the \ci\ emission is in the upper layers, it points to an additional UV source in the midplane. This requires a powerful source of high-energy photons, which could be an accretion disk around a protoplanet.

\section{Calculation of the \ci\ Column Density}\label{Appendix2}

We use the peak intensity to calculate the \ci\ column density in the kinematic anomaly, in order to estimate the mass flow. From radiative transfer equations for continuum subtracted line emission, the observed intensity is given by the following:

\begin{equation}\label{Eq:RT1}
    J_{\nu_0 , \mathrm{obs}} = J_{\nu_0}(T_{\rm exc}) \Big( 1 - \mathrm{exp}(-\tau_{\nu_{\rm peak}}) \Big) ,
\end{equation}

\noindent where $J_{\nu_0}$ is the mean   intensity at the line peak,  $T_{\rm exc}$ the excitation temperature, which we set to 45 K, and $\tau_{\nu_{\rm peak}}$ the optical depth at the line peak. 

The optical depth at the line peak is given by:

\begin{equation}\label{Eq:tau}
\tau_{\nu_{\rm peak}} = \frac{c^2}{8\pi\nu^2}\Big(\mathrm{exp}\Big(\frac{h\nu}{k_BT} \Big) -1 \Big) A_{ul}\phi_{\nu_{\rm peak}} N_u,
\end{equation}

\noindent with $c$ the speed of light, $h$ the Planck's constant, $k_B$ the Boltzmann's constant, $A_{ul}$ the Einstein coefficient for the respective line, $\phi_{\nu_{\rm peak}}$ the value of the line profile at the peak of the line, and $N_u$ the column density of the targeted species at the excited state.

The value of the line profile at the peak is given by:
\begin{equation}\label{Eq:lineprof}
\phi_{\nu_{\rm peak}} =\frac{1}{\sqrt{2\pi}\sigma},
\end{equation}

\noindent with:

\begin{equation}\label{Eq:lineprof2}
\sigma^2 = \frac{c^2}{2\nu^2 \Big( \frac{2k_BT}{m_{CI} } +v_{\rm turb}^2  \Big)},
\end{equation}

\noindent where $m_{CI}$ is the mass of \ci\ , $v_{\rm turb}$ a turbulent velocity component, which we set to 200 m s$^{-1}$, and $T$ the temperature.

Assuming $J_{\nu} = 4\pi B_{\nu}$, the blackbody emission, and combining it with Eqs. \ref{Eq:RT1}, \ref{Eq:tau} and  \ref{Eq:lineprof}; the excited column density of \ci\ is recovered. To estimate the total \ci\ column density, we used the partition function as follows:

\begin{equation}\label{Eq:NCI}
N_u = N_{CI} \frac{g_u}{Q}\exp{\Big(\frac{-E_u}{k_B T_{\rm exc}}\Big )},
\end{equation}

\noindent with  $g_u$ the degeneracy of the upper level, $Q$ the partition function, $E_u$ the energy of the upper level,  and $T_{\rm exc}$ the excitation temperature.  Above 30 K, the brightness of the \ci\ $^3$P$_1$-$^3$P$_0$ line is relatively insensitive to temperature.  Excitation temperatures between 40 K and 100 K produce column densities, N$_{\rm CI}$, in the range 3.2-3.9 10$^{17}$ cm$^{-2}$, i.e, only a 22\% change.  Thus, adopting a value of 45 K for the column density retrieval does not produce significant changes in the column density given the other sources of uncertainty. All the specific constant values for the \ci\ $^3$P$_1$-$^3$P$_0$ line are taken from the Leiden Atomic and Molecular Database, LAMDA \citep{Lamda}.

\section{Velocity Centroid in different Datasets}\label{datasets}

\begin{figure}[ht!]
\plottwo{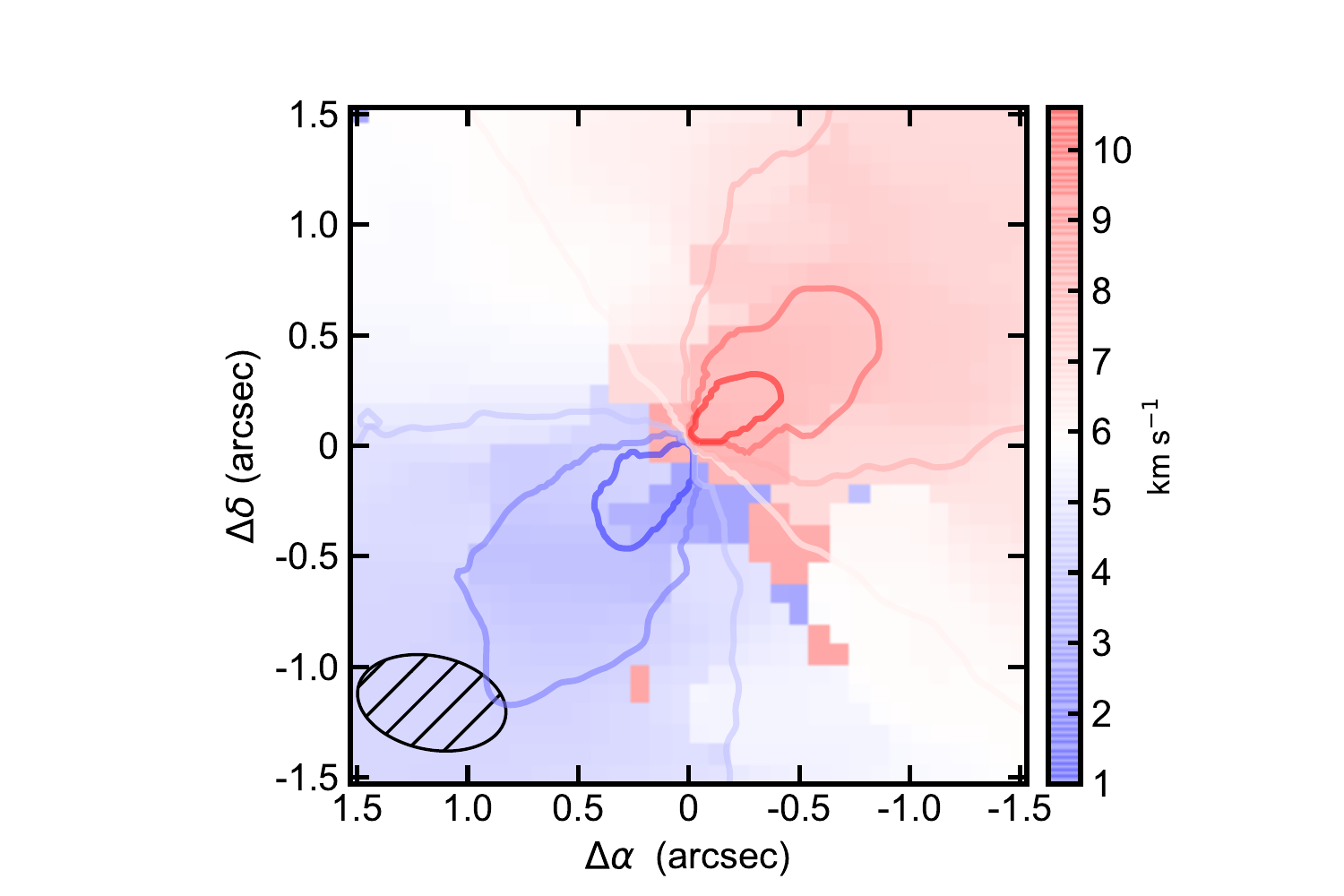}{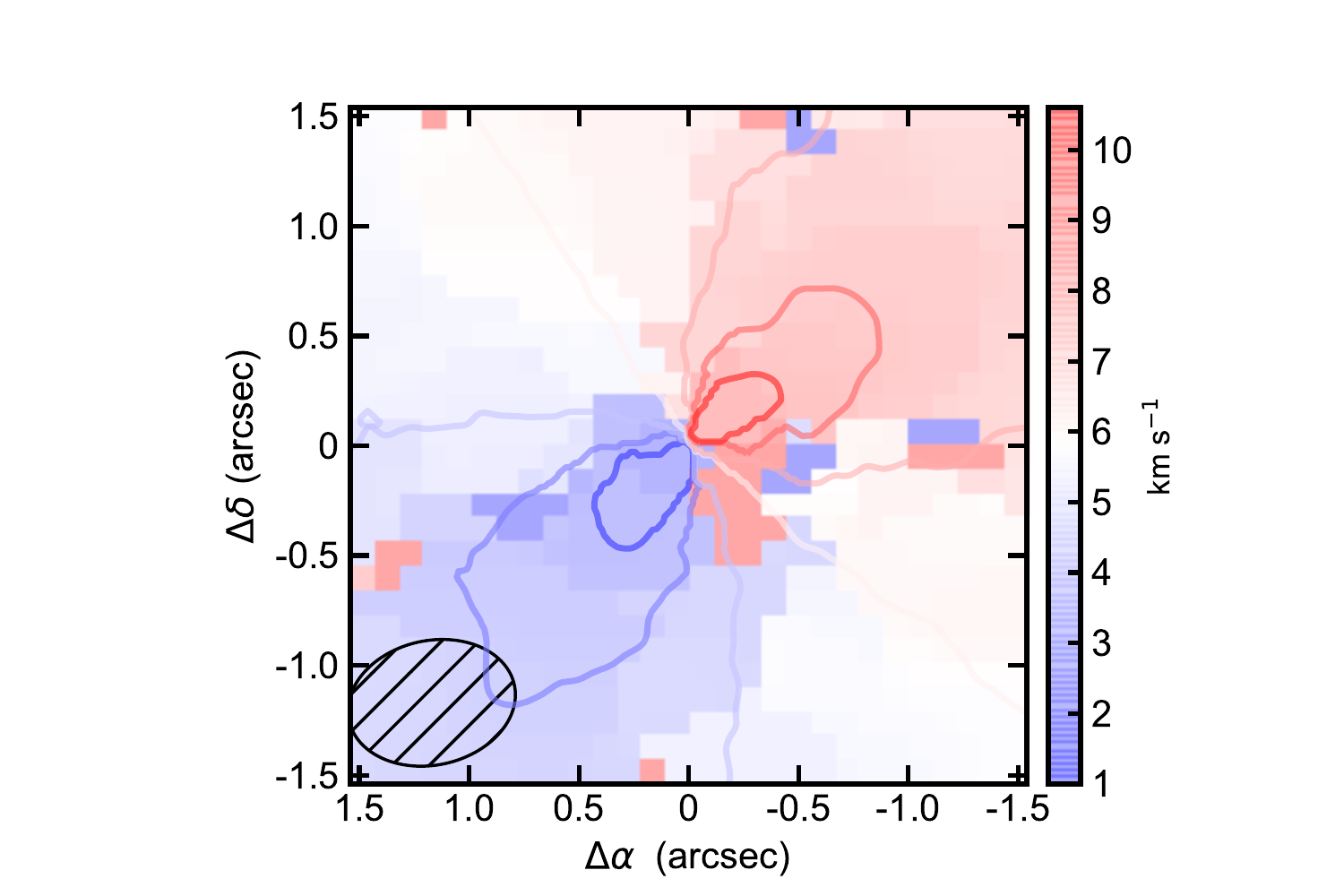}

\caption{Velocity centroid map of the \ci\ $^3$P$_1$-$^3$P$_0$ line for the   2013.1.00527.S and 2015.1.01137.S. datasets on the left and right respectively. The kinematic feature in the SW of the disk is observed in both datasets independently.
\label{fig:CO_vs}}
\end{figure}

For completeness, we also show the velocity centroids in Figure \ref{fig:CO_vs}. Both datasets show the kinematic feature, although the statistical significance of the feature for each dataset independently is not high enough to assure the presence of the feature. 

\bibliography{CI_paper.bib}
\bibliographystyle{aasjournal}



\end{document}